\documentclass[twocolumn,showpacs,preprintnumbers,amsmath,amssymb]{revtex4}

\usepackage{graphicx}
\usepackage{amssymb}
\usepackage{dcolumn}
\usepackage{bm}
\usepackage{upgreek,tabularx}
\usepackage{color}

\begin{document}


\title{Influence of rhombohedral stacking order in the electrical resistance
of bulk and mesoscopic graphite}

\author{M. Zoraghi, J. Barzola-Quiquia, M. Stiller, A. Setzer and P. Esquinazi}
\affiliation{Division of Superconductivity and Magnetism, Institute for Experimental Physics II, University of Leipzig,
D-04103 Leipzig, Germany}

\author{G. H. Kloess~ and T. Muenster} \affiliation{Institut f\"ur Mineralogie,
Kristallographie und Materialwissenschaft, Fakult\"at f\"ur Chemie
und Mineralogie, Universit\"{a}t Leipzig, Scharnhorststra{\ss}e
20, D-04275 Leipzig, Germany}

\author{T. L\"uhmann and I. Estrela-Lopis}
\affiliation{Institute for Medicine Physics and Biophysics,
University of Leipzig, D-04107 Leipzig, Germany}





\begin{abstract}
  The electrical, in-plane resistance as a function of temperature $R(T)$ of bulk and mesoscopic
  thin graphite flakes obtained from the same batch was investigated. Samples thicker than
  $\sim 30$~nm show metalliclike contribution in a temperature range that increases
  with the sample thickness, whereas
a  semiconductinglike behavior was observed for thinner samples.
The temperature dependence of the in-plane
resistance of all measured samples and several others
from literature can be very well
explained between 2~K and 1100~K  assuming three contributions in parallel: a metalliclike
  conducting path at the interfaces between  crystalline regions,
composed of two semiconducting
  phases, i.e.~Bernal and
  rhombohedral stacking. From the fits of  $R(T)$ we obtain a semiconducting
  energy gap of $110 \pm 20$~meV for the rhombohedral and $38\pm 8~$meV for the Bernal phase.
  The presence of these crystalline phases
  was confirmed by x-ray diffraction
  measurements. We review similar experimental data from literature of the last 33 years and  two
  more theoretical models used to fit $R(T)$.

\end{abstract}

\maketitle

\section{Introduction}
\label{intro}

 Graphite, a layered material built by  weakly coupled graphene
sheets, is a material being studied experimentally and
theoretically for more than fifty years. Usually, the graphene
layers adopt an hexagonal $ABAB\dots$ (2H) stacking sequence
called the Bernal~\cite{BERN} structure. However, another stable
phase in graphite has  an $ABCABC\dots$ stacking order called
rhombohedral (3R) graphite~\cite{LIPS}. According to  early
literature \cite{LIPS,kelly}, high quality graphite samples can be
composed of up to $\sim 30\%$ rhombohedral and the rest Bernal
phase. Recent studies on exfoliated few-layer graphene (FLG) using
Raman spectroscopy, show domains of different stacking order with
$\sim 15~\%$ of the total area displaying 3R stacking~\cite{LUI}.
The domains exhibiting 3R stacking in FLG are stable after annealing to
$800~^\circ$C~\cite{LUI} and to $1000~^\circ$C in bulk
graphite~\cite{MATU,FREIS}.

Wallace calculated the band structure of graphite within
tight-binding approximation with the result that graphite behaves
like a semiconductor with a vanishing small energy
gap~\cite{WAL1,WAL2}.
 The existence of
a finite energy gap in graphite was proposed by Mrozowsky based on
the analysis of electrical resistivity and diamagnetism of
polycrystalline graphite~\cite{MRO1,MRO2}. Recent theoretical work
on 3R graphite suggests the formation of an energy gap that should
become smaller increasing the number of graphene layers of the 3R
phase~\cite{GUIN,KOSH1,KOSH2}. By means of angle-resolved
photoemission~\cite{COLE}  the existence of an energy gap of the
order of 100~meV in trilayer 3R graphite  was obtained. The study
of the electrical transport properties of bulk graphite dates back
to Dutta for single crystals \cite{DUTTA} and to Reynolds for
natural and polycrystalline samples \cite{REYN}. Since then, none
of the large number of published studies on the transport
properties of graphite considered the influence of the 3R phase,
even though there is no doubt about its existence in usual
graphite samples.

The presence of the 3R phase in graphite samples can have a
further, drastic influence on their transport properties. Recent
theoretical work predicts a topological protected flat band at the
surface of 3R graphite~\cite{kop11}, which was recently confirmed
experimentally~\cite{PIER}. Moreover, assuming a finite
Cooper-pair coupling, this flat band might trigger
high-temperature superconductivity~\cite{HEIK1,HEIK2,kop13}, which
should exist also at the embedded interfaces between Bernal and 3R
crystalline phases~\cite{kopbook,mun13}  and/or at twisted single
Bernal phases~\cite{esqarx14}. Therefore, the presence of both
stacking orders in a graphite sample can have clear competitive
contributions to the conductivity of real samples.

In this work, the influence of 3R stacking on the in-plane
resistance of 11 samples, bulk and mesoscopic graphite  flakes,
was studied. The presence of both stacking orders was confirmed by
x-ray diffraction (XRD), see Section~\ref{exp}. In Section~\ref{res} we show that
assuming the contributions of the two
crystalline stacking plus a metalliclike contribution from the
interfaces in parallel, we are able to fit the temperature
dependence of the resistance of all samples in a broad temperature
range. From the fits we obtain an energy gap for the 3R stacking
regions in our samples in  agreement with results from
literature~\cite{COLE}. Furthermore, in Section \ref{com} we show that our model describes
with high degree of accuracy published resistance data of macroscopic and
mesoscopic samples of different thickness in a broad range of temperature, i.e. from~2~K to 1100~K \cite{end83,oha97,gut09}, using as fitting parameters similar energy gaps.
A detailed comparison of the proposed models
in literature to explain the temperature
dependence of the resistance of graphite \cite{oha97,gut09} in a large temperature
range, reveals that the  proposed models do not really fit
the published data,
independently of the used fitting parameters.

\section{Experimental details and sample quality characterization}
\label{exp}
The graphite samples used for experiments  were obtained from
highly oriented pyrolytic graphite (HOPG) bulk material from
Advanced Ceramic with a rocking curve width of $0.4^{\rm o}$ and
metallic impurities in the ppm range~\cite{spe14}. Investigated
mesoscopic flakes were produced on top of silicon substrates caped
with a 150~nm thick insulating silicon nitride ($\rm Si_3N_4$).
For this work, the flakes were produced by a rubbing method
already described in previous publications~\cite{JBQ1}. After
selecting suitable samples, electron beam lithography was used to
print the structures for the electrodes, which were sputtered with
a bilayer of~Cr$/$Au with a thickness of $\approx5$~nm and
$\approx30$~nm, respectively. The temperature dependent resistance
of the samples was measured in a commercial $\rm ^4He$ cryostat,
within the temperature range of 2~K to 310~K.  Low noise
resistance measurements were performed using an AC Bridge (Linear
Research LR-700), with a constant current $\lesssim 5~\upmu\rm A$.

\begin{figure}
\includegraphics[width=1\columnwidth]{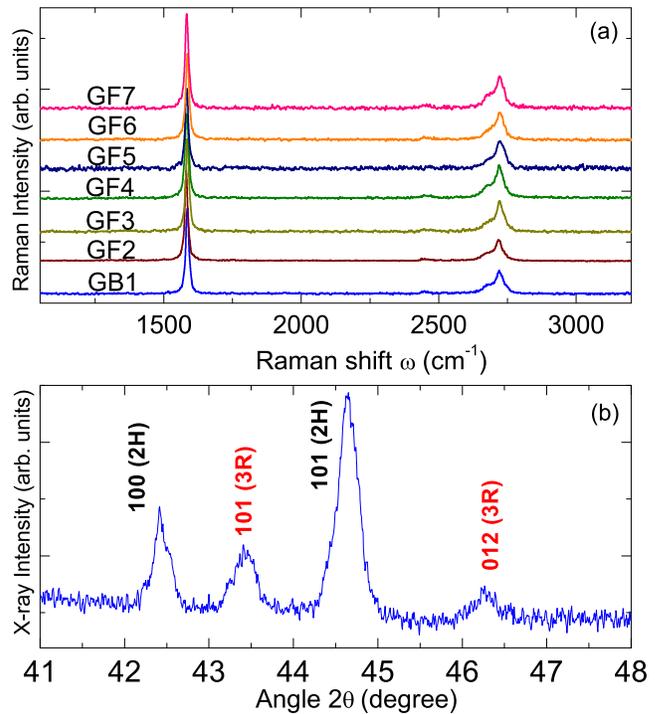}
\caption{\label{fig:Fig3} (a) Raman spectrum of some of the
investigated samples. GB1 is the bulk sample, see Table~I. (b)
x-ray diffraction pattern for the bulk sample in a restricted
angle region. The labels with the Miller indices  near the Bragg
peaks indicate whether the maxima belong to Bernal (2H) or
rhombohedral (3R) phase.}
\end{figure}


\begin{table*}[htb]
\begin{tabularx}{\textwidth}{@{}p{35pt}p{65pt}XXXXXXXX@{}}
\hline\hline
Sample &   Thickness (nm) & $ a_{1}$ & $ a_{2}$ & $E_{g1}$ (meV) & $E_{g2}$ (meV) & $E_a$ (meV)&$R_0$&$R_1$&$R_2$\\
\hline
GB1      & 6000 & 1.2 E-4 &10 E-4   &   106 & 40 & 4.0&0.33&0.0018&0.61\\
GF2      & 85 &  8.2 E-5 &7.3 E-4 &  97  & 25 & 4.9&0.28&0.0028&1.14\\
GF3      & 50 &  9.1 E-5 &12.6 E-4 &  104   & 35 & 3.9&0.6&0.0012&0.68\\
GF4      & 115 &  5.8 E-5 &7.8 E-4 &  90  & 29 & 4.8&0.6&0.0075&0.77\\
GF5      & 80 &  7.7 E-5 &7.0 E-4 &   107 & 39 & 4.1&0.72&0.0016&0.67\\
GF6      & 95 &  9.7 E-5 &9.5 E-4 &  98  & 32 & 4.5&0.73&0.0011&0.72\\
GF7      & 80 &  8.5 E-5 &10 E-4 &  105 & 36 & 4.0&0.73&0.0011&0.69\\
GF8      & 72 &  8.1 E-5 &8.9 E-4 &  113 & 42 & 2.9&0.93&0.001&0.39\\
GF9    & 35 &  7.9 E-5 &6.6 E-4 &  124 & 51 & 3.8&1.3&0.0003&0.11\\
GF10     & 35 &  9.1 E-5 &9.8 E-4 &  103 & 43 & --&1.98&-0.0012&-\\
GF11     &57 & 6.1 E-5 & 6 E-4 &  114 & 44& 3.5&0.52&0.0027&0.45\\
\hline\hline
\end{tabularx}
\caption{Summary of samples, their thickness and different
parameters obtained from the fits of $R(T)$ to Eq~(5). $E_{g1}$
corresponds to the rhombohedral phase, $E_{g2}$ to the Bernal one.
The unitless coefficients $a_{1,2}$ are the normalized
corresponding prefactors of the semiconducting contributions in
Eq.~(\ref{eq:Rtot2}) for the Bernal $(a_2)$ and rhombohedral $(a_1$) phases.
The unitless coefficients $R_{0,1,2}$ are the corresponding normalized
  parameters of the metallic-like contribution in Eq.~(\ref{eq:Rtot2}) from the interfaces.}
\label{tab:t1}
\end{table*}


The structural quality of all samples was investigated by Raman
and XRD measurements. For this purpose, a confocal micro-Raman
microscope was used (alpha~$300+$, WITec) with an incident laser
light with $\lambda = 532$ nm and a maximal power of $3$~mW. The
Raman results of some selected samples are shown
in~Fig.~\ref{fig:Fig3}(a). The most intense peaks in the Raman
spectra of graphene and graphite are expected at $\simeq 1580~{\rm
cm}^{-1}$ (the $G$-peak) and at $\simeq 2700~\rm cm^{-1}$ (the
$G'$ peak)~\cite{FERR1}.  The so-called
 $D$-peak at
$\simeq 1350~{\rm cm}^{-1}$ is related to the disorder present in
the material~\cite{FERR2,FERR4}. The Raman results indicate,  see
Fig.~\ref{fig:Fig3}(a), that all samples have the same structural
order as the initial bulk material (sample GB1) with no evidence
of disorder within experimental resolution.

The XRD measurements were done using a Bruker D8 Discover
 (Cu K$^\alpha$ radiation at
40~kV and 40~mA) with a GADDS-detector system (V{\AA}NTEC-500).
Note that there are several peaks not suitable for distinguishing
both stacking modifications of graphite. Both the $(00l)$ and
$(hh0)$ of the 2H and 3R stacking are superposed. Therefore, the
$2\Theta$ range $40-47^\circ$ was selected to determine and
approximately quantify the rhombohedral phase in the samples, see
Fig.~\ref{fig:Fig3}(b), because the reflexes are not superposed in
this particular range. Note, however, the reflex intensities are
$\lesssim 1\%$ of the maximum relative intensity. The Rietveld
refinement using TOPAS 4.2 results in $86 \pm 3$~wt.\% for the 2H
and $14 \pm 3~$wt.\% for the 3R phase. We conclude that the
presence of 3R stacking in our samples is confirmed. Scanning
transmission electron microscopy (STEM) measurements for similar
samples show that they are composed of many crystalline regions
with aligned $c$-axis but with different $a-b$-axes orientations,
i.e.~twisted Bernal crystalline regions~\cite{JBQ1}. As XRD
results indicate, some of the crystalline regions have 3R
stacking. Between both regions, Bernal and 3R as well as between
twisted Bernal regions, interfaces are formed.
Following STEM pictures \cite{JBQ1,gar12} and electron back
scattering diffraction (EBSD) done on similar samples \cite{gon07,gar08}
the crystalline
regions have in general a lateral size of the order of tenths of micrometer
and a thickness varying from a few nanometers to $\sim
400~$nm. Because the percentage of 3R phase in our samples remains
below 20\% according the XRD results,  we expect that probably the thin crystalline
regions in our samples \cite{JBQ1,gar12} have the 3R phase.

\section{Own experimental data and proposed model}
\label{res}

\begin{figure}
\includegraphics[width=1\columnwidth]{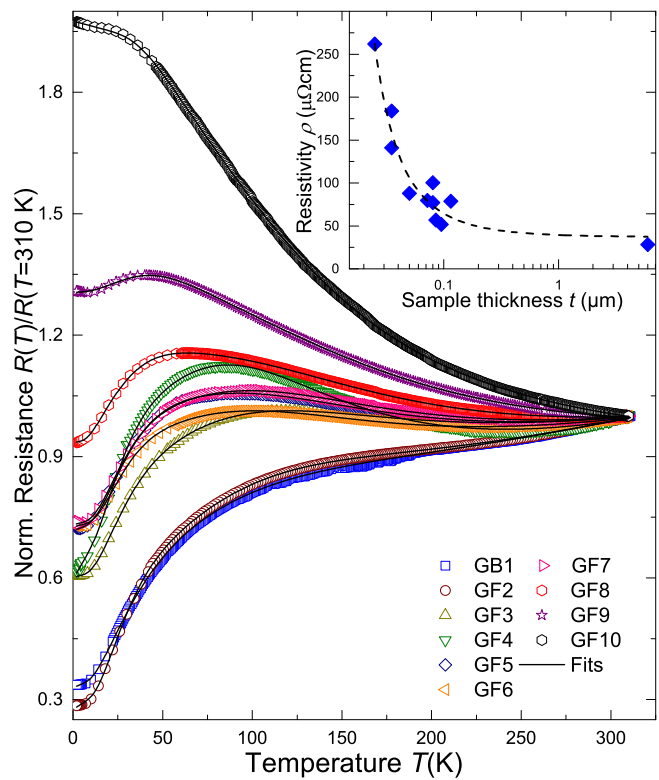}
\caption{\label{fig:Fig1} Normalized resistance results of all
  investigated samples. The lines are fits of the data to Eq.~(\ref{eq:Rtot2}) with
  the parameters of Table~I. The inset shows the calculated resistivity
  as a function of the thickness. The dashed line is a
  guide to the eye.}
\end{figure}

The results of the temperature dependence of the normalized
resistance of all investigated samples are presented
in~Fig.~\ref{fig:Fig1}. Although all samples were obtained from
the same initial material, different temperature dependence can be
observed. Some samples show a metalliclike behavior in all
temperature range, such as the GB1 (bulk), or for the multilayer
graphene (MLG) flakes GF2 and GF3.  Some samples exhibit a
combination of metallic and semiconductinglike behavior (e.g. GF4
and GF5). The samples GF9 and GF10  show only semiconductinglike
behavior with a saturation at low temperatures. The overall
results are in agreement with those published earlier~\cite{JBQ1}.
At first, considering that the samples were produced using the
same initial material and that the structural quality between
samples does not appear to differ, the results shown in
Fig.~\ref{fig:Fig1} are not obvious. Considering the
internal structure of the initial bulk sample revealed by
STEM~\cite{JBQ1}, it is clear that the significant difference is
given by the sample thickness. A dependence of the calculated
resistivity on the sample thickness is shown in the inset of
Fig.~~\ref{fig:Fig1}. The resistivity increases with decreasing
thickness, indicating that the resistivity in relatively thick
graphite samples is not constant and that these samples should not
be considered as  homogeneous material. In other words, the calculated
resistivity using the sample geometry and resistance data should not be
taken as an intrinsic value of the graphite structure. Therefore, in what follows we discuss and
show resistance, not resistivity data.

The usual thickness dependence of the resistivity in metalliclike
systems, such as Cu~\cite{KE} or Ag~\cite{TAN}, can be described
using the theory of Fuchs-Sondheimer~\cite{FUC,SOND}, which
considers the influence of scattering processes at the sample
surface. However, this does not apply for such an anisotropic material
as graphite, which consists of very weakly coupled stacked layers of 2D graphene sheets,
with each single graphene sheet already conducting.
For the same reason and the lack of evidence of internal disorder within the MLG samples according to
the Raman results, see Fig.~\ref{fig:Fig3}, the sign change of temperature coefficient of resistance
in thin graphite samples ($\partial R/\partial T < 0)$ cannot  be accounted for by enhanced electron scattering rate  in
analogy with the Mooij rule used to interpret the electrical conduction of disordered transition metal alloys \cite{moo73}.

The decrease of the
resistivity with increasing thickness can be interpreted as a
consequence of the increasing metalliclike contribution of the
interfaces between crystalline regions \cite{JBQ1,gar12}. We note
that the formation of metalliclike regions at interfaces is not a
new concept and was already observed in many oxide materials where
even superconductivity was found at very low
temperatures~\cite{REY1,REY2}.
Garc\'ia~\textit{et~al.}~\cite{gar12} proposed a simple,
phenomenological  model to understand the $R(T)$, which consists
of two main contributions in parallel, one is originated from the
interfaces $(R_{\rm i}(T))$ and the other from the crystalline,
semiconducting regions $(R_{\rm s}(T))$.

There should be no doubt that if the XRD results indicate that two
well defined  stacking orders exist in our samples, and STEM
pictures show also clearly different crystalline regions
\cite{JBQ1,gar12,esqarx14}, one should take care that
two-dimensional (2D) boundaries exist between those crystalline
structures, embedded in the graphite sample. Moreover, 2D
interfaces can occur also between twisted layers of graphene,
which are characterized by a rotation angle and lateral
translation. This type of 2D interfaces produces the so-called
moir\'e patterns in the electron density of states, found on the
surface of a macroscopic HOPG sample already in 1990~\cite{kuw90}
and supported by several, recently done studies (for a recently
published review see \cite{chap7}). Our XRD results as well as
other evidence \cite{chap7} indicate that one cannot be sure that
a transport property like the electrical resistance provides an
intrinsic property of ideal graphite of the region checked between the
voltage electrodes. Therefore, as a first guess we write the total
measured resistance of the graphite samples as the sum of two
parallel resistances formulated as \cite{gar12}:
\begin{equation}\label{eq:Rtot1}
R(T)^{-1}=R_{\rm i}^{-1}(T)+R_{\rm s}^{-1}(T)\,.
\end{equation}

\begin{figure}
\includegraphics[width=0.9\columnwidth]{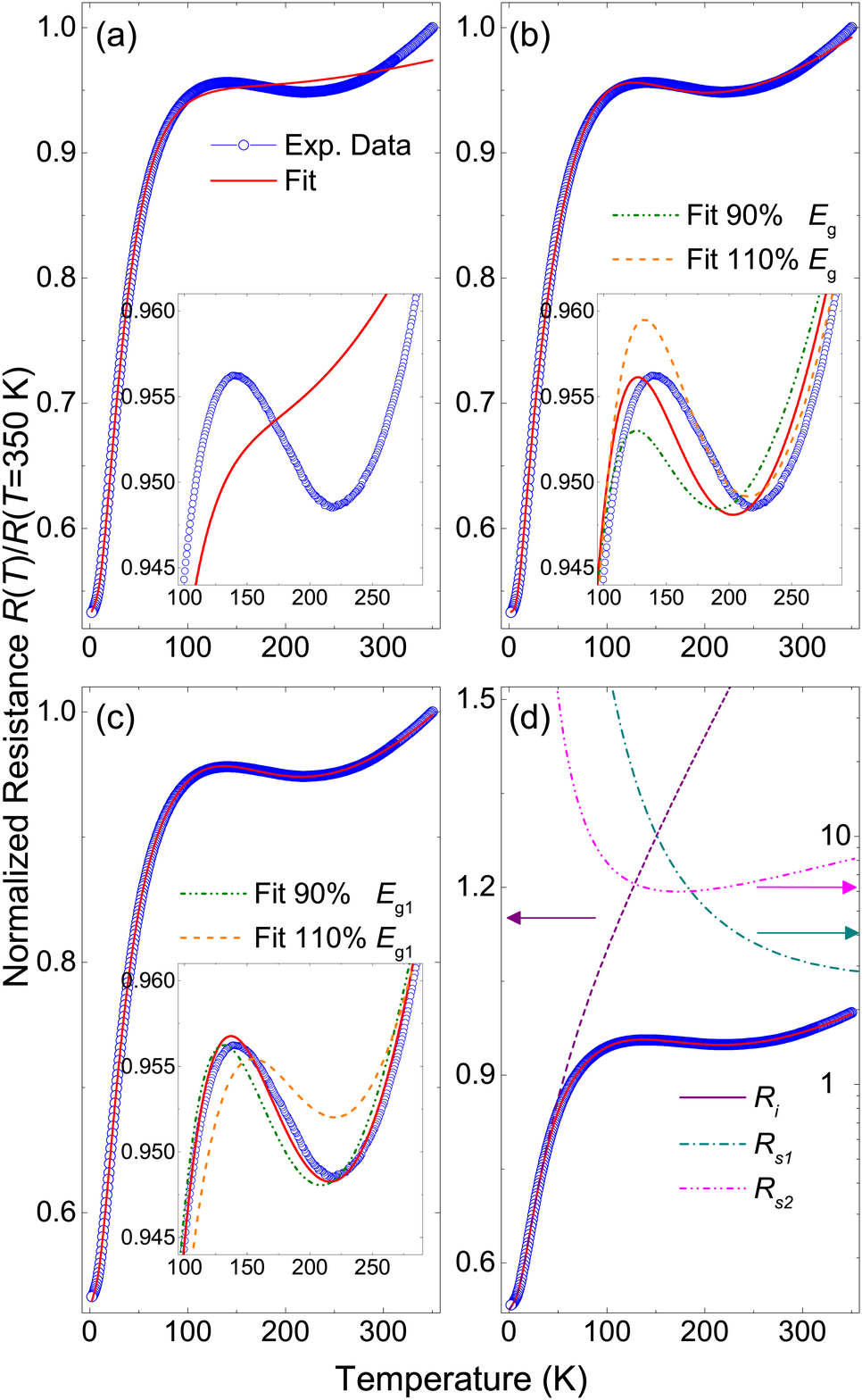}
\caption{\label{fig:Fig2} $R(T)$ of sample GF11.  The fits are
shown as
  continuous lines. In (a) the data were fitted using
  Eqs.~(\ref{eq:Rtot1},\ref{eq:Rif},\ref{eq:Rsc})
  with $a(T)/R(350)=$1.46, a semiconducting energy gap
$E_g = 26~$meV, an activation energy $E_a = 3.8~$meV, and the
parameters $R_{0,1,2}/R(350)= 0.53, 0.002, 0.57$, respectively.
(b) The same data and a fit to similar equations as in (a) but
with $a(T) = a_0 T^{3/2}$. The parameters of the best fit (red
curve) are:
   $a_0/R(350)=1.7 \times 10^{-4},
E_g = 89~$meV, $E_a = 2.95~$meV, $R_{0,1,2}/R(350)= 0.53,
6.5\times 10^{-4}, 0.54$. The other curves are obtained fixing the
$E_g$ value 10\% above and below the best fit value and leaving
all other parameters free. In (c) and (d) the data were fitted
using
  Eq.~(\ref{eq:Rtot2}) (red curve)
  with the parameters given in Table~I.
  The insets in (a-c)
  expands the data and  fits at
  high temperatures. The other fit curves were obtained as in (b) changing
   the value of $E_{g1}$ by $\pm 10$\%. In (d) each in-parallel contribution to
the resistance is shown separately.
 A similar fitting procedure is obtained for all samples.
 See as further example the one for sample GF3 in Fig.~\ref{GF3}.}
\end{figure}

\begin{figure}
\includegraphics[width=0.9\columnwidth]{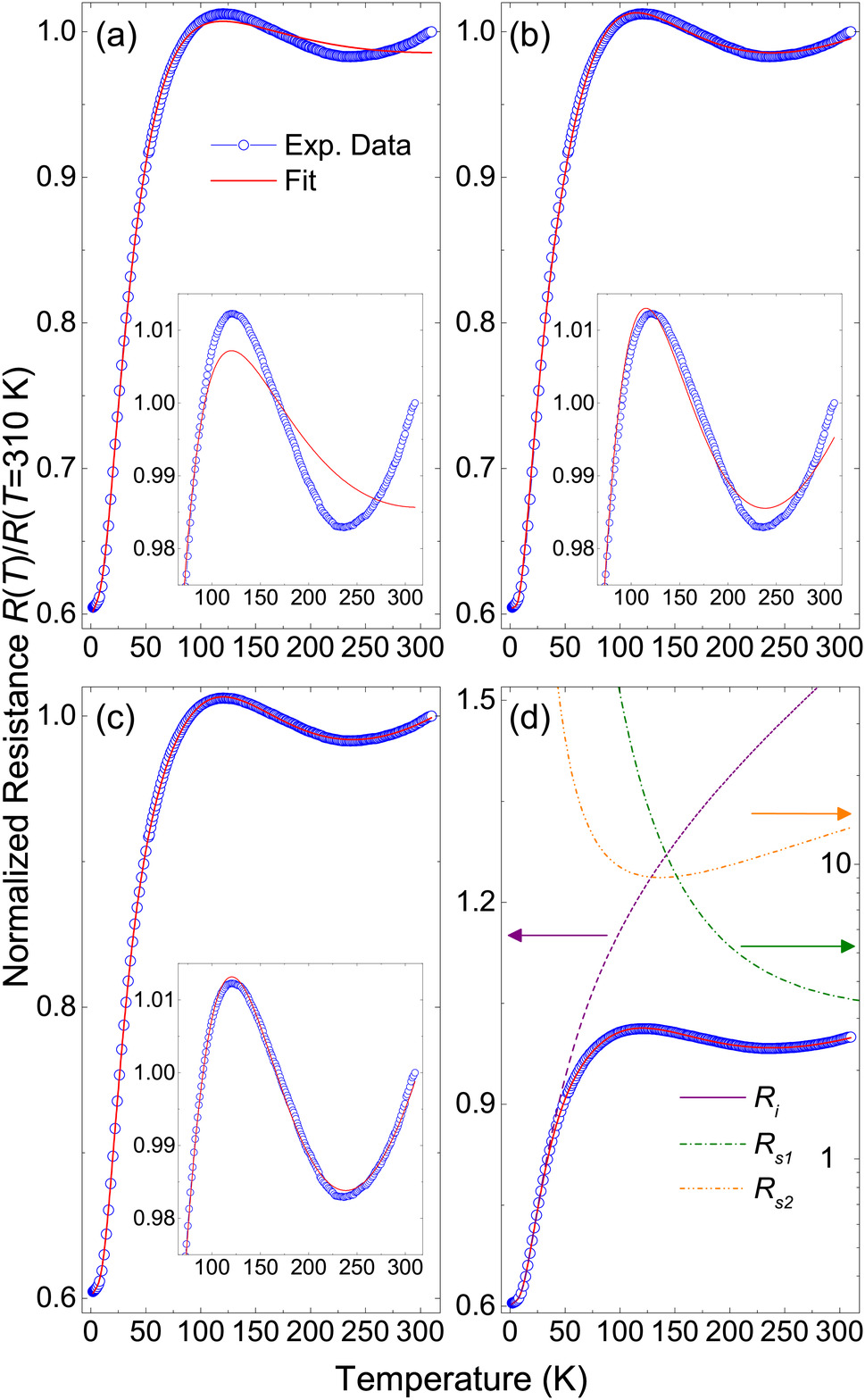}
\caption{\label{GF3} $R(T)$ of sample GF3.  The fits are shown as
  continuous lines. In (a) the data were fitted using
  Eqs.~(1,2,3) with $a(T)/R(310)=1.38$,
  a semiconducting energy gap
$E_g = 29.5~$meV and an activation energy $E_a = 3.85~$meV, and
the parameters $R_{0,1,2}/R(310)= 0.59, 0.0018, 0.58$,
respectively. (b) The same data and a fit to similar equations as
in (a) but with $a(T) = a_0 T^{3/2}$. The parameters of the best
fit (red curve) are:   $a_0/R(310)=1.92 \times 10^{-4}, E_g =
87~$meV, $E_a = 3.2~$meV, $R_{0,1,2}/R(310)= 0.60, 2.9\times
10^{-4}, 0.59$.  In (c) and (d) the data were fitted using
  Eq.~(5) (red curve)
  with the parameters given in Table~I.
  The inset in (a-c)
  expands the data and  fits at
  high temperatures. In (d) each in-parallel contribution to
the resistance is shown separately.}
\end{figure}

To obtain a good fit at low temperatures, where the metalliclike
behavior overwhelms, it is necessary to assume an interface
contribution to the total resistance of the form:
\begin{equation}\label{eq:Rif}
R_{\rm i}(T)=R_0+R_1T+R_2\exp\left( \frac{-E_a}{k_{\rm B}T} \right),
\end{equation}
where the coefficients $R_0$, $R_1$, $R_2$ as well as the
activation energy $E_a$ are free parameters. The temperature
independent term $R_0$ represents the residual resistance at low
temperatures. Note that this residual, temperature independent
resistance is necessary to assume in the metalliclike
contribution to
Eq.~(\ref{eq:Rtot1}), especially due to its influence at low
temperatures. A similar residual, in series term is not necessary to assume
in the semiconducting contribution. The linear term contribution (usually much weaker
than the exponential one, i.e. $R_1 \ll R_{0,2}$) is expected to
come from the longitudinal acoustic (LA) phonon
scattering~\cite{PIET,STAU,PARK}, for example.  This contribution
was already observed in graphene samples produced on ${\rm
SiO}_2$~\cite{HAO} and in suspended graphene~\cite{BOLO}. A
similar, exponential third term of Eq.~(\ref{eq:Rif}) was already
used to describe the temperature dependence of two dimensional
electron-hole systems formed at the interfaces of non-conducting
materials, e.g.~GaAs/AlAs heterostructures~\cite{HANE} or $p$-type
SiGe~\cite{COLER}.  This thermally activated contribution was
first used by Kopelevich \textit{et al.}~\cite{yaknar} and later by
Takumoto \textit{et al.}~\cite{TAKU} to fit the metalliclike increase in
the resistance with temperature observed in bulk graphite samples.
The origin of this contribution, however, remains still
controversial. For example, it could be explained using a
percolation of electron-hole liquid~\cite{HE}, disorder and
electron-electron interactions~\cite{FINKE}, or through the
enhanced spin-orbit interaction by broken inversion
symmetry~\cite{PUDA}. However, the probable origin of this
thermally activated term could be related to the superconductivity
localized at the interfaces between Bernal and 3R stacking. We
note that a similar exponential dependence has been observed in
granular Al-Ge~\cite{sha83} for a particular Al concentration.
This thermally activated behavior can be understood on the basis
of the Langer-Ambegaokar-McCumber-Halperin (LAMH)
model~\cite{lan67,mcc70} that applies to narrow superconducting
channels in which thermal fluctuations can cause phase slips. The
value of the activation energy $E_a \sim 4~$meV obtained from the
fits remains similar for all measured samples, see Table~I.

The second term in Eq.~(\ref{eq:Rtot1}) is related to the
crystalline parts, which we assume to behave as intrinsic
semiconductors, given by:
\begin{equation}\label{eq:Rsc}
R_{s}(T)=a(T)\cdot{\rm exp}\left( \frac{+E_{\rm g}}{2k_{\rm B}T} \right),
\end{equation}
where $a(T)$ is a mobility-dependent prefactor, $E_{\rm g}$ is the
semiconducting energy gap and $k_{\rm B}$ is the Boltzmann
constant. The temperature dependent parameter $a(T)$ can be
written as:
\begin{equation}\label{eq:aT}
a(T)=\frac{1}{N (\mu_h+\mu_e) T^{3/2}},
\end{equation}
where $N$ is a temperature independent constant and $\mu_e$ and
$\mu_h$ are the electron and hole mobilities.  In the work of
Garc\'ia~{\it \textit{et al.}}~\cite{gar12}, $a(T)$ was assumed to be
constant, which results of assuming $\mu_{h,e}\propto T^{-3/2}$,
and it is a good approximation for typical
semiconductors~\cite{NORT,CANA}.

As example, we show the fitting procedure to Eq.~(\ref{eq:Rtot1})
for sample GF11, see Fig.~\ref{fig:Fig2}(a). The parameters
obtained from the fittings have a maximum standard deviation of
$\sim 10$\% and  low correlation effects. Therefore, we restrict
ourselves to show the best fits obtained after a careful test of
the correlation effects between the free parameters. The assumed
metalliclike contribution given by Eq.~(\ref{eq:Rif}) fits well at
low temperatures. However, in the inset we can clearly see that at
higher temperatures, the fit considerably differs from the
experimental data.  We assume that this deviation at high
temperatures is partially a consequence of the approximation that
$a(T)$ is constant. Therefore, we improve the model assuming that
$\mu_{e,h}\propto T^{-3}$ following experimental studies in
graphite flakes~\cite{gar12,ESQ1}. Taking this into account, we
obtain $a(T)=a_0\cdot T^{3/2}$ and include it in
Eq.~(\ref{eq:Rsc}). Using this  in Eq.~(\ref{eq:Rtot1}) we can fit
the experimental data at high temperatures better than before (see
Fig.~\ref{fig:Fig2}(b) and its inset). In the inset we show the
change of the best possible fits changing manually 10\% $E_g$ and
leaving all other parameters free.

Similar or even worse results for the other investigated samples
are obtained assuming a single semiconducting contribution. From
our XRD data we know that graphite is composed of two phases,
Bernal and 3R stacking, and therefore we now consider these two
independent and in-parallel contributions in our model to describe
the temperature dependence of the resistance. For this purpose,
Eq.~(\ref{eq:Rtot1}) is modified by adding a new semiconducting
contribution in parallel:
\begin{equation}\label{eq:Rtot2}
R_{\rm t}(T)^{-1}=R_{\rm i}^{-1}(T)+R_{\rm s1}^{-1}(T)+R_{\rm s2}^{-1}(T),
\end{equation}
where $R_{\rm s1}$ and $R_{\rm s2}$ correspond to 3R and Bernal
stacking. Using this new assumption, we can very well fit the
experimental data over all temperature range for all samples, see
Fig.~\ref{fig:Fig2}(c) and Fig.~\ref{fig:Fig1}. The contributions
of each component of Eq.~(\ref{eq:Rtot2}) are plotted in
Fig.~\ref{fig:Fig2}(d) as lines together with the data of sample
GF11. Further example of these fits can be seen for sample GF3 in
Fig.~\ref{GF3}.

\begin{figure}
\includegraphics[width=0.9\columnwidth]{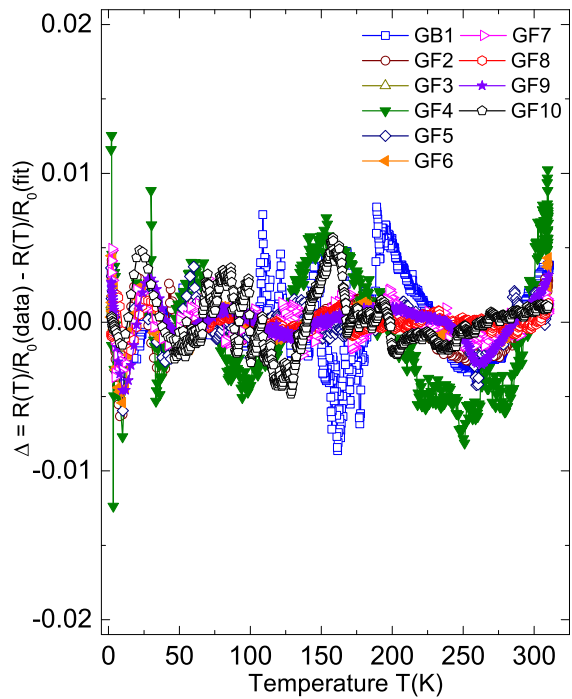}
\caption{\label{del} The difference between the measured
normalized resistance $R(T)/R(T_0)$  and the fitted curve
 vs.~temperature, with the parameters from Table~\ref{tab:t1}. A
difference of 0.01 means a deviation of the order of 1\% or smaller. $T_0$ is
an arbitrary selected temperature, e.g. 310~K in this case.}
\end{figure}

Both, the $R(T)$ of bulk and thin MLG samples, can be very well
fitted to Eq.~(\ref{eq:Rtot2}) with the parameters listed in
Table~I. To show the accuracy of the fit of the data to
Eq.~(\ref{eq:Rtot2}) we plot in Fig.~\ref{del} the difference
between the data and the fits defined as
\begin{equation}
\Delta = \frac{R(T)}{R(T_0)}(\textrm{data}) -
\frac{R(T)}{R(T_0)}(\textrm{fit})\,.
\label{Delta}
\end{equation}
The results in Fig.~\ref{del} indicate that the used model in this
work predicts the measured normalized resistance with an accuracy
better than 1\% in the whole temperature range. Note that there is
no systematic deviation, i.e. $\Delta$ fluctuates around zero in
the whole temperature range.

One may  argue that the excellent fits to
the experimental data cannot be taken too seriously because of the large number
of free parameters, see Table I. However, a quick look at the values of the parameters
obtained from the fits of such different temperature dependences of $R(T)$ (see Fig.~\ref{fig:Fig1}),
indicates the following interesting facts that relativize
to some extent that argument:\\
(1) Regarding the interface metalliclike contribution given by Eq.~(\ref{eq:Rif}), we note that it
defines mainly the behaviour of $R(T)$ at temperatures $T \lesssim 100~$K, if a metalliclike contribution
is present. The main part of the increase of $R(T)$ with temperature  (Eq.~(\ref{eq:Rif})) follows
always a thermally activated contribution
with an activation energy of the order of 4~meV.\\
(2) Although comparatively small,  the introduction of
a linear in temperature term with the prefactor $R_1$ appears necessary.
For several samples, upon their thickness, a pure
metalliclike behavior ($\partial R/\partial T > 0$)  between 100~K and 300~K is not obvious,  neither  a
semiconductinglike one. It turns out that
the data showing pronounced maxima and minima cannot be well fitted without $R_1$. Independently of the possible
justification based on electron-phonon interaction \cite{PIET,STAU,PARK}, this term is necessary also to fit $R(T)$ of other carbon-based materials like nanocrystal graphite thin films \cite{cho14}.\\
(3) In case the metalliclike interfaces contribution is negligible, e.g. sample GF10 in Fig.~\ref{fig:Fig1}, a good fit of $R(T)$ below $\sim 70$~K is only possible if we, as expected, neglect the thermally activated contribution $R_2 = 0$ and
$R_1 < 0$. The negative sign of $R_1$ is not expected if electron-phonon interaction would play a role.
However, such a negative, nearly linear in $T$ term in $R(T)$ especially at low temperatures,
has been observed in nano-Ag grains \cite{yus00} as well as in ion-beam-deposited W, Pd and Pt
nanostructures \cite{bar09}. Its origin appears to be induced  by
 interfaces with very low order  or disordered structures
at  the interfaces between the Ag, PdC, WC and PtC nanograins. It
is appealing to suggest that the origin of this small, negative
$R_1 T$ term is related the disordered interfaces, i.e. the one to
vacuum and the one at the substrate of the FLG sample. On the
other hand, the best fit of $R(T)$ for sample GF10 is obtained
assuming still the existence of the two stacking orders, see Table
I. In this case it is possible that not metallic but disordered
interfaces still exist in the 35~nm thick sample with a resistance
too high to play a main role in $R(T)$ at high temperatures. Thus,
the contribution of these disordered interfaces can be seen only
at low enough temperatures, in contrast to samples where the
metalliclike interfaces dominate. Thinner samples may have only
one of the semiconducting contributions, as shown in
\cite{gar12} for a 13~nm thick HOPG sample.\\
(4) The behavior at $T > 300$~K is given mainly by the semiconducting parts given by $R_{s1}$ and $R_{s2}$ in parallel.\\
(5) The values of the semiconducting energy gaps are similar for all samples with a ratio
$E_{g1}/E_{g2} = 2.9 \pm 0.3 $.\\
(6) The weight ratio between  the two semiconducting contributions given by  $a_1/a_2 = 0.10 \pm 0.02$, a value of the
order of the mass ratio between the 3R and 2H phases in our samples obtained from XRD, see Section II.

The values obtained for the activation energy $E_a$ are similar
to those from literature~\cite{yaknar,TAKU,gar12}
and are, compared to 2DEG systems, one order of magnitude larger.
In the work of Garc\'ia~ {\it \textit{et al.}}~\cite{gar12} the samples were
investigated to $T\approx275$~K and the obtained semiconducting
energy gap $\sim 40~$meV was attributed to the main phase of the
sample, the Bernal stacking. Similar small band gaps have been
observed in rhombohedral Bi~\cite{LIU} and $\rm
Bi_{0.88}Sb_{0.12}$ alloy~\cite{SAUN}. The values of $E_{g1} \sim
100~$meV obtained from the fits of $R(T)$ for all samples are in
good agreement with that of ARPES \cite{COLE}. Therefore, the
energy gaps obtained from the fitting process can be related to
the two semiconducting phases.

\section{Comparison of our model with published resistance data and other theoretical
models from literature }
\label{com}

\begin{figure}
\includegraphics[width=1\columnwidth]{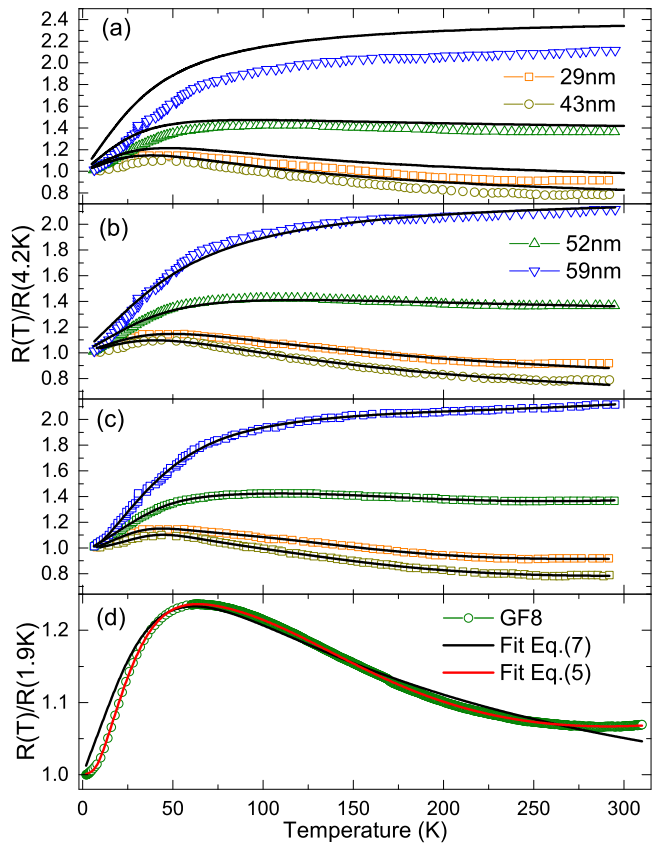}
\caption{\label{fig:oha1} (a-c): Normalized resistance vs. temperature of the data from \cite{oha97} of the samples with thickness between 29~nm and 59~nm. In (a) we show the fit curves following Eq.~(\ref{eq:oh}) with the parameters given in that publication, see also Figs.~\ref{fig:oha2}(b,c). (b) Similar to (a) but with the best fits with the parameters to Eq.~(\ref{eq:oh}) shown in Figs.~\ref{fig:oha2}(b,c) (blue circles). (c) Similar to (a,b) but with the fits to Eq.~(\ref{eq:Rtot2}) and with the parameters given in Table II. (d) Normalized resistance vs. temperature for sample GF8 and the best fits of the data to Eq.~(\ref{eq:oh}) and Eq.~(\ref{eq:Rtot2}).}
\end{figure}

\begin{figure}
\includegraphics[width=1\columnwidth]{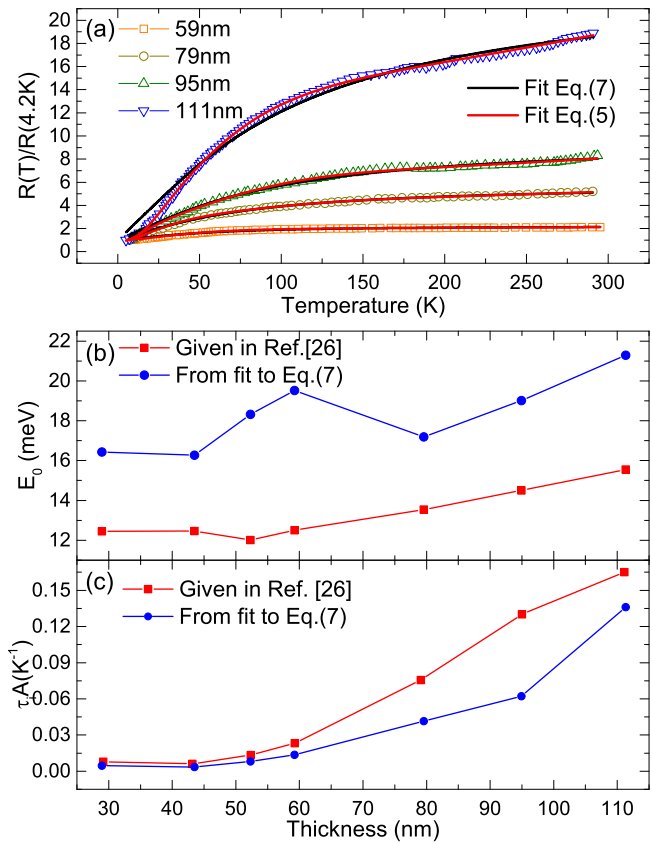}
\caption{\label{fig:oha2} (a) Normalized resistance vs. temperature of the data from \cite{oha97} of the samples with thickness between 59~nm and 111~nm and the best fits to Eq.~(\ref{eq:oh}) and Eq.~(\ref{eq:Rtot2}). (b) Band overlap energy $E_0$ vs. thickness obtained from the fits to the data of \cite{oha97}: red squares are the parameters of the original publication, the blue circles the parameter obtained from the best fits to Eq.~(\ref{eq:oh}). (c) Similar to (b) but for the scattering relaxation time parameter $\tau_iA$ obtained from the fits of the data of \cite{oha97}.}
\end{figure}

The effects of the thickness of  graphite samples on the
electrical properties were already studied by Ohashi \textit{et al.}
\cite{oha97}  by cleaving a kish graphite sample with a relatively
large rest resistance ratio of 32. Our results for $R(T)$ and its
thickness dependence are basically similar to those from
\cite{oha97}  and \cite{JBQ1}, compare our results in
Fig.~\ref{fig:Fig1} and those in Figs.~\ref{fig:oha1} and
\ref{fig:oha2}(a), i.e.  the smaller the thickness of the sample
the lower is the temperature where a metalliclike behavior is
observed below 300~K. The model used by Ohashi \textit{et al.} to interpret
the obtained data is based on a two-band model and a theory for
lattice vibration in thin-carbon films that includes
electron-Rayleigh-wave interaction \cite{sug88}. The main
assumptions of the model are: - three dimensional graphite is a
semimetal because the valence band overlaps slightly the
conduction band, - the degree of the overlap of these two bands
depends on the film thickness and is included in the model by the
free parameter $E_0$, and - two relaxation rates, one due to
lattice defects $\tau_i^{-1}$ and the other due to lattice
vibrations proportional to temperature $AT$ \cite{sug88}, included
in the model as the free parameter $\tau_iA$. According to this
model the normalized resistance is given by the expression:
\begin{equation}
 \frac{R(T)}{R(T_0)}=\frac{E_0}{2k_{\rm B}T
 \ln\left(1+\exp\frac{E_0}{k_{\rm
 B}T}\right)}\left(1+\tau_iAT\right)\,.
 \label{eq:oh}
\end{equation}
The experimental data from \cite{oha97} for samples of thickness
between 111~nm and 29~nm are shown in Figs.~\ref{fig:oha1} and
\ref{fig:oha2}(a). To check the accuracy of the fits to
Eq.~(\ref{eq:oh}) to the authors data we show in
Fig.~\ref{fig:oha1}(a) the data of samples with thickness between
59~nm and 29~nm taken from \cite{oha97}. Using the same parameters
from that publication one realizes that the fit to
Eq.~(\ref{eq:oh}) is bad, see Fig.~\ref{fig:oha1}(a).
Nevertheless, we left the two parameters $E_0$ and $\tau_iA$ free
and tried to get the best fits of the experimental data to
Eq.~(\ref{eq:oh}). The results of these fits are shown in
Figs.~\ref{fig:oha1}(b) and \ref{fig:oha2}(a) and the obtained
free parameters as a function of thickness are shown in
Figs.~\ref{fig:oha2}(b) and \ref{fig:oha2}(c), together with those from
the original publication \cite{oha97}.

\begin{figure}
\includegraphics[width=1\columnwidth]{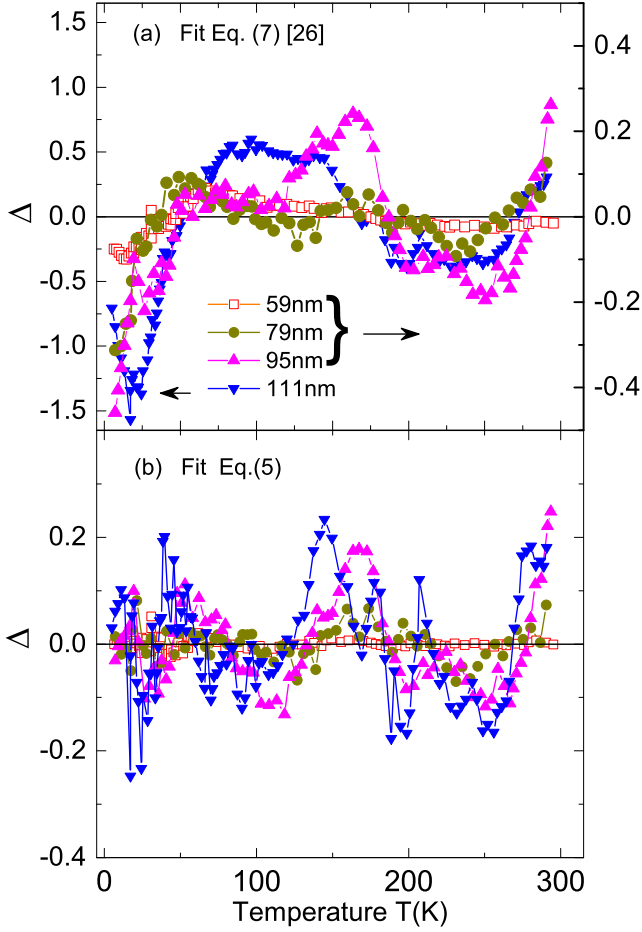}
\caption{\label{D-O} Difference $\Delta$ between  the data and: (a)  the
fits using Eq.~(\ref{eq:oh}) from Ohashi \textit{et al.} \cite{oha97}, but with the best fit parameters (shown
as blue circles in Fig.~\ref{fig:oha2}(b,c)).  (b) Similar difference but from the fits done following  our model
given by Eq.~(\ref{eq:Rtot2}) and with the parameters given in Table II.}
\end{figure}

We note that the overlapping energy $E_0$ does
show a non monotonous change with thickness, in contrast to the
authors conclusion, with a maximum at a thickness around 60~nm,
see Fig.~\ref{fig:oha2}(b). Within the assumptions of the model
this behavior is not expected and it is difficult to provide any
simple explanation, unless the samples with thickness between
50~nm and 60~nm would have had some peculiarities (defects,~etc.)
that there other samples do not. To check this
speculation and also the accuracy of the fits of the data of
\cite{oha97} to Eqs.~(\ref{eq:Rtot2}) and (\ref{eq:oh}) we show: -
in Figs.~\ref{fig:oha1}(c) and \ref{fig:oha2}(a) the fits to the
Ohashi \textit{\textit{et al.}} data \cite{oha97} with our model, - the best fit of
the data of our sample GF8 to their Eq.~(\ref{eq:oh}) (and to our
Eq.~(\ref{eq:Rtot2})) in Fig.~\ref{fig:oha1}(d), - and in
Fig.~\ref{D-O} the difference $\Delta$ between  the data and: (a)  the
fits using Eq.~(\ref{eq:oh}) from Ohashi \textit{et al.} but with the best fit parameters (shown
as blue circles in Fig.~\ref{fig:oha2}(b,c))  and (b) the fits according to our model
given by Eq.~(\ref{eq:Rtot2}). It is
clear that the Ohashi \textit{et al.} model does show systematic and much larger deviations
from the experimental data (more that 100\% in certain temperature range) than with
our model.
The parameters obtained from the fits
of Ohashi \textit{et al.} data to our model given by Eq.~(\ref{eq:Rtot2})
are given in Table~II. It is interesting to note that the obtained
parameters are similar to those from the fits to our data, see Table I.

\begin{table*}\label{tab2}
\begin{tabularx}{\textwidth}{@{}p{55pt}p{65pt}XXXXXXXX@{}}
\hline \hline
Sample from&   Thickness (nm) & $ a_{1}$ & $ a_{2}$ & $E_{g1}$ (meV) & $E_{g2}$ (meV) & $E_a$ (meV)&$R_0$&$R_1$&$R_2$\\
\hline
\cite{end83}&$\dag$&8.2E-5&5.6E-4&127&38&5.7&0.96&0.004&1.55\\
\cite{oha97} &29&4.5E-5&4.6E-4&107&37&9.8&0.96&0.0059&0.17\\
\cite{oha97}&43&3.4E-5&2.4E-4&109&33&12.2&0.98&0.003&6.29\\
\cite{oha97}&52&6.5E-5&4.7E-4&124&47&4&0.96&0.008&0.09\\
\cite{oha97}&59&1E-4&5.5E-4&154&61&3.7&0.97&0.0069&0.79\\
\cite{oha97}&79&2.9E-4&1.9E-3&116&35&2.7&0.66&0.0416&0.86\\
\cite{oha97}&95&5.4E-4&3.2E-3&104&42&6.5&0.69&0.0363&2.23\\
\cite{oha97}&111&1.6E-3&6.9E-3&104&42&6.5&0.71&0.0486&13.8\\
\cite{gut09}&$\ast$&2.9E-4&2.5E-3&107&29&6.7&0.75&0.0371&1.91\\
\hline \hline
\end{tabularx}
\caption{Best fit parameters to Eq.~(\ref{eq:Rtot2}) of the experimental data of the electrical
  resistance vs. temperature from Endo \textit{et al.}~\cite{end83}, Ohashi \textit{et al.}~\cite{oha97}  and Gutman \textit{et al.}~\cite{gut09}, including
  the samples thickness.  $E_{g1}$ corresponds to the energy gap of the
  rhombohedral phase, $E_{g2}$ to the Bernal one. The unitless coefficients $a_{1,2}$ are the corresponding normalized
  prefactors of the semiconducting contributions in Eq.~(\ref{eq:Rtot2}) for the Bernal ($a_2$) and rhombohedral ($a_1$) phases.
  The unitless coefficients $R_{0,1,2}$ are the corresponding normalized
  parameters of the metallic-like contribution in Eq.~(\ref{eq:Rtot2}) from the interfaces, similar to the ones shown in Table~\ref{tab:t1}.
  ($\dag$): The sample
  was a pristine benzene-derived fiber heat treated to $2900^\circ$C with the graphite crystalline structure according to the authors \cite{end83}.  There is no
  information on that publication on the thickness of the sample, but it is written that a special gold paste
  was used to form the electrical contacts between the sample
  and lead wires. This suggests that sample was macroscopic, i.e. probably of mm size. ($\ast$): In Ref.~\cite{gut09} there is no information on
  the size of the measured HOPG sample grade A. However, because the contacts between the wires and the sample were made with using
  silver or graphite paint, we believe also that the sample was macroscopic, not mesoscopic.}
\end{table*}

\begin{figure}
\includegraphics[width=1\columnwidth]{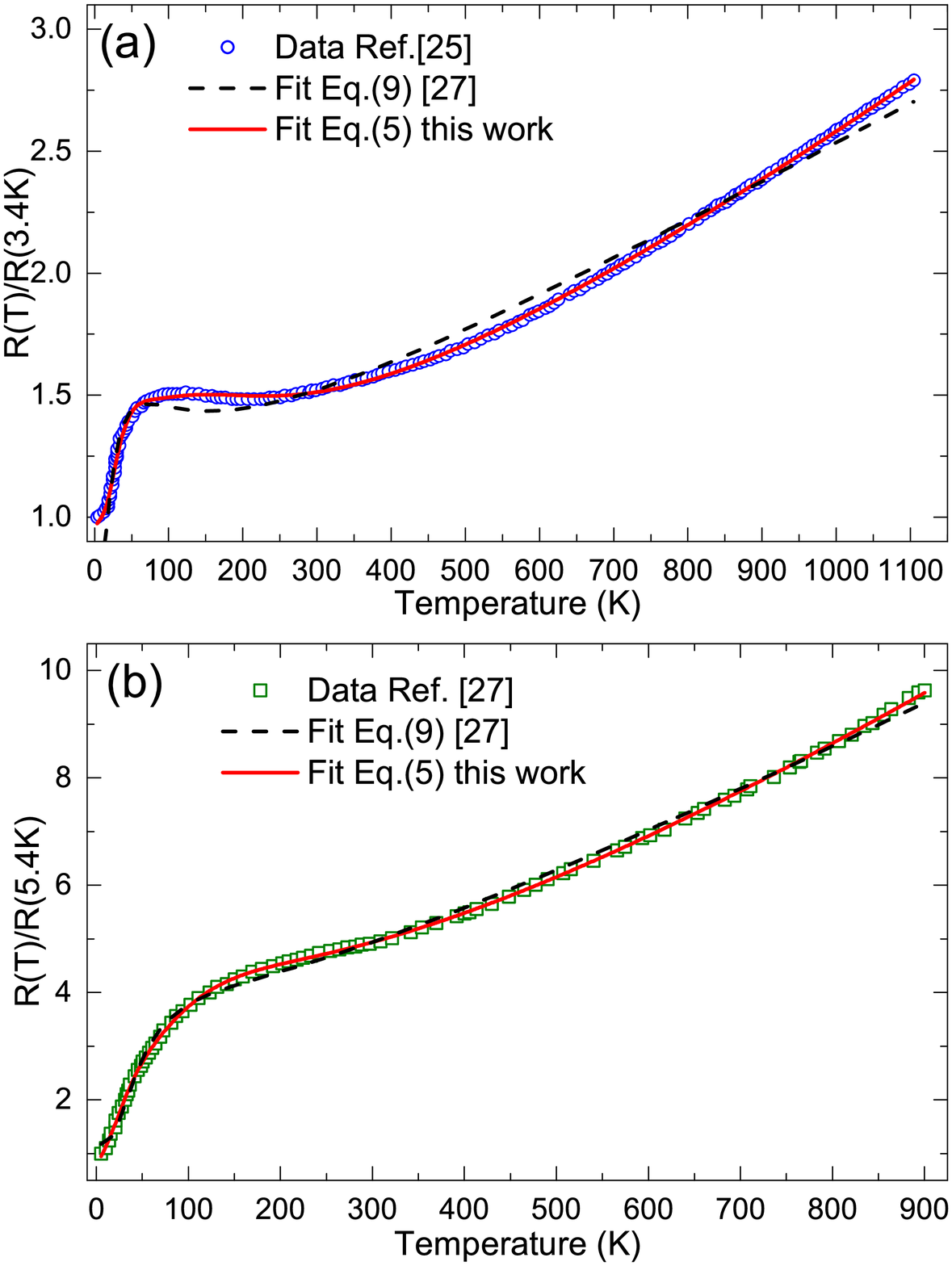}
\caption{\label{REG} Normalized resistance vs. temperature of: (a) Endo \textit{et al.} \cite{end83} and (b) Gutman \textit{et al.} \cite{gut09}. The dashed lines are the best fits of the data to Eq.~(\ref{gut2}) and the continuous (red) lines
are those to Eq.~(\ref{eq:Rtot2}), see Table II.}
\end{figure}

\begin{figure}
\includegraphics[width=1\columnwidth]{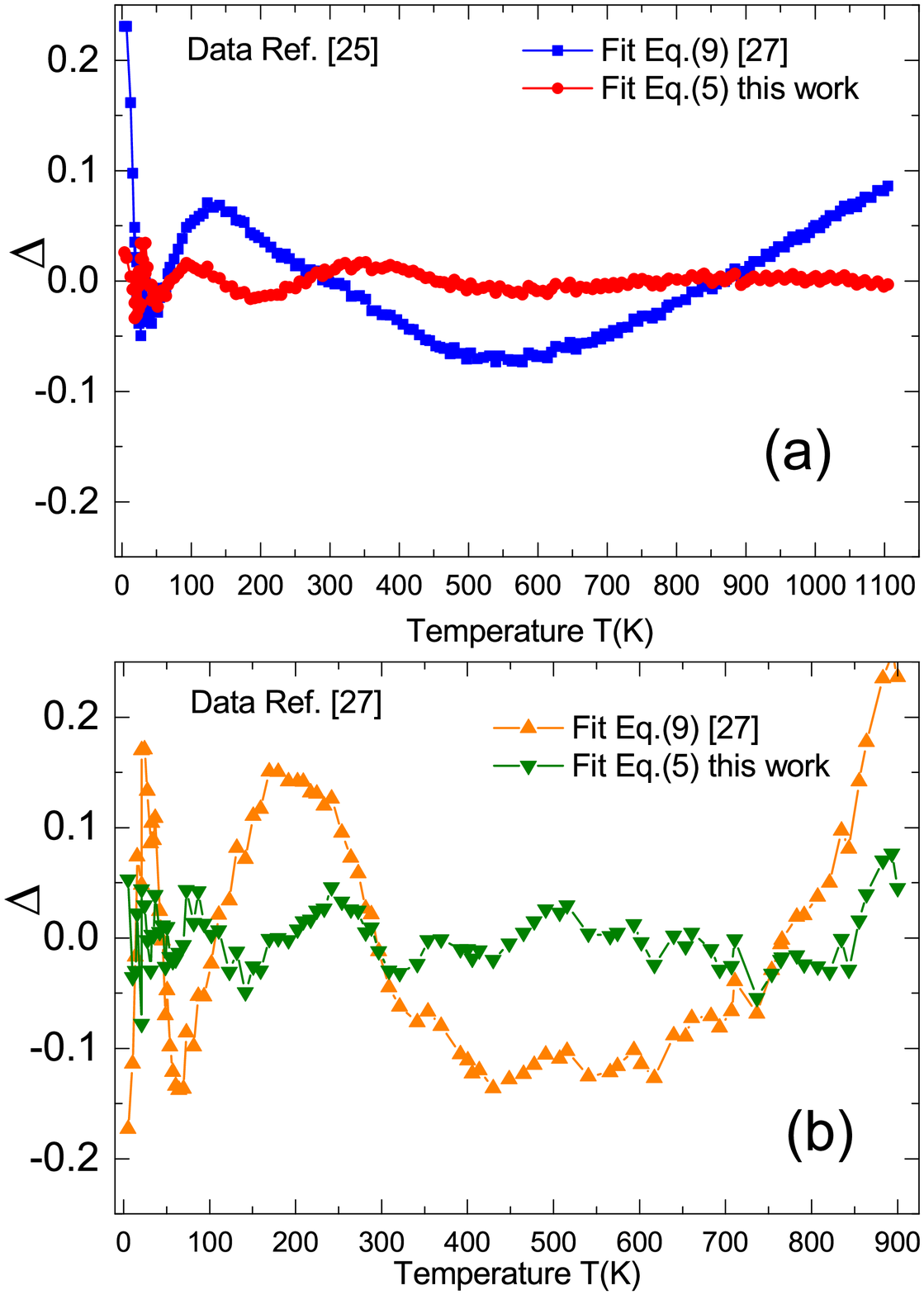}
\caption{\label{DEG} Difference $\Delta$, defined in Eq.~(\ref{Delta}), for the data of (a) Endo \textit{et al.} \cite{end83} and the Eqs.~(\ref{gut2}) and (\ref{eq:Rtot2}) and similarly for the data of (b) Gutman \textit{et al.} \cite{gut09}.}
\end{figure}

In what follows we compare one further model published by Gutman \textit{et al.} \cite{gut09} and our model with in-plane resistance data obtained from bulk graphite samples up to 1100~K. Figure \ref{REG} shows the experimental data of the normalized resistance vs. temperature obtained from  (a) \cite{end83}
and (b) \cite{gut09}. It is interesting to note that the resistance of both samples increases with temperature above 300~K in a similar way, although according to the authors in \cite{gut09}, one expects a compensation between the increase in the number of carriers and the decrease in the scattering time, i.e. a saturation of the resistance. Therefore,  an extra intervalley scattering  of charge carriers by high-frequency, graphene-like optical phonons was assumed in \cite{gut09} that provides according to those authors the necessary increase of the resistance with temperature.
According to the model in \cite{gut09} the resistance is given by the expression:
\begin{equation}
\varrho=\frac{c}{e^2}\left(\frac{1}{\tau_0}+\alpha T\right)\frac{1}{\epsilon^*}+\frac{c}{e^2}\frac{1}{a_0T\bar\tau}\exp\left(-\frac{\omega_0}{T}\right)\,,
\label{gut}
\end{equation}
where the first term accounts for the low temperature behavior ($\tau_0^{-1}$ is the scattering rate due to impurities and $\alpha T$ due to soft phonons, similarly as in Eq.~({eq:oh}) and $\epsilon^* \sim E_F$, the Fermi energy. The second term is due to intervalley scattering. Equation~(\ref{gut}) has four free parameters, $\tau_0, \alpha, \bar\tau$, and $\omega_0$, being the last two the effective electron-phonon relaxation time and the frequency of the longitudinal optical mode ($E_{2g}$) at the $\Gamma$ point. To fit  the data of Endo \textit{et al.} \cite{end83} and the data of Gutman \textit{et al.} \cite{gut09} with this model we have used the normalized resistance following Eq.~(\ref{gut}) as:
\begin{equation}
\frac{R(T)}{R(T_0)}=P_1 + P_2 T+(P_3/T) \exp(-P_4/T)\,.
\label{gut2}
\end{equation}
For the  fit shown in Fig.~\ref{REG}(a) of the data of Endo \textit{et al.} \cite{end83} we obtained as best fit parameters: $P_1=0.764, P_2=0.00169~$K$^{-1}$, $P_3=91.7~$K and $P_4=\omega_0=56.8$~K=4.89~meV. in Fig.~\ref{REG}(b) we show the fit of the data of Gutman \textit{et al.}  to Eq.~(\ref{gut2}) using the same values for the free parameters as in the original publication \cite{gut09}. As comparison, we show in the same figures the fits of the data to our Eq.~(\ref{eq:Rtot2}) with the
parameters shown in Table II. For a  better recognition of the differences between experimental data and fits, Fig.~\ref{DEG} shows the difference $\Delta$ for both data using Eq.~(\ref{gut2}) from \cite{gut09} and our Eq.~(\ref{eq:Rtot2}). In this figure it is clearly observed that our model fits the resistance temperature dependence in the whole temperature range to 1100~K with a remarkable accuracy, better than 5\% (relative) and without any systematic deviation from the main experimental temperature behaviour, in contrast to the model given by Eq.~(\ref{gut2}). From all these results we may conclude that the increase of $R(T)$ in graphite is due to the temperature increase expected for a small-gap
semiconducting material  with a mobility that decreases  with temperature as $T^{-3}$.

\section{Conclusion}
Concluding, we have investigated the longitudinal resistance of a
bulk and a series of mesoscopic graphite samples obtained from the
same initial material with similar structural quality. Our results
show that the transport properties of bulk graphite are not
unique, as they depend strongly upon the amount of interfaces
present in the material. By fitting the temperature dependence of
the resistance we found indications for the contribution of the
semiconducting rhombohedral phase with an energy gap similar to
the one reported in literature. XRD measurements reveal the
presence of the rhombohedral and Bernal phases in the graphite
material used in our experiments. From our interpretation we can conclude that
the metalliclike contribution to the electrical resistance is not intrinsic of
ideal graphite but due to
 interfaces between, e.g., Bernal and 3R stacking.

 Independently of the fit parameters used,
  none of the published models can fit $R(T)$ as accurate  as the one proposed in this study.
  The available data from literature and in a broad temperature range indicate also the
  existence of the two stacking orders with similar energy gaps as the samples studied in
  this work.

We note that these interfaces
might be the reason for the superconductinglike behavior at very
high temperatures observed in the magnetization of bulk and
treated graphite powders \cite{yakovjltp00,sch12,schcar}, in the
transport properties of TEM graphite lamellae where a direct
contact to the interfaces has been achieved \cite{bal13}, as well
as in stapled graphite flakes \cite{kaw13}.\\

Acknowledgments: We acknowledge fruitful discussions with T.
Heikkil\"a, G.~Volovik and Y. Kopelevich.


\end{document}